\documentclass[%reprint,
 amsmath,amssymb,aps,twocolumn,a4paper,%fleqn
]{revtex4-1}

\usepackage{graphicx}% Include figure files
\usepackage{dcolumn}% Align table columns on decimal point
\usepackage{bm}% bold math
\usepackage{lipsum}

\newcommand{\mv}[1]{\ensuremath{\mathbf{#1}}} % for vectors
\newcommand{\gv}[1]{\ensuremath{\mbox{\boldmath$ #1 $}}} 
\newcommand{\avg}[1]{\left \langle #1 \right \rangle} % for average
\usepackage[colorlinks]{hyperref}  
\usepackage{cleveref}
\hypersetup{linkcolor=blue,citecolor=blue,urlcolor=blue}

\usepackage{titlesec}
\titleformat{\section}
  {\centering\normalfont\fontsize{11}{15}\bfseries}{\thesection}{1em}{}

\usepackage[toc]{appendix}

\crefname{figure}{fig.}{figs.}
\Crefname{figure}{Figures}{Figures}

\begin{document}

\setcounter{page}{1} %first page number

\title{The Influence of Particle Softness on Active Glassy Dynamics}

\author{Vincent E. Debets$^{1,2}$, Liesbeth M.C. Janssen$^{1,2*}$}

\affiliation{$^{1}$Department of Applied Physics, Eindhoven University of Technology, P.O. Box 513,
5600 MB Eindhoven, The Netherlands\\$^{2}$Institute for Complex
Molecular Systems, Eindhoven University of Technology, P.O. Box 513,
5600 MB Eindhoven, The Netherlands\\}
\email{l.m.c.janssen@tue.nl}

\begin{abstract}%
{\noindent Active matter studies are increasingly geared towards the high-density or glassy limit. This is mainly inspired by the remarkable resemblance between active glassy materials and conventional passive glassy matter. Interestingly, within this limit it has recently been shown that the relaxation dynamics of active quasi-hard spheres is non-monotonic and most enhanced by activity when the intrinsic active length scale (e.g., the persistence length) is equal to the cage length, i.e.\ the length scale of local particle caging. This optimal enhancement effect is claimed to result from the most efficient scanning of local particle cages. Here we demonstrate that this effect and its physical explanation are fully retained for softer active spheres. We perform extensive simulations of athermal active Brownian particles (ABPs) and show that the non-monotonic change of the relaxation dynamics remains qualitatively similar for varying softness. We explain quantitative differences by relating them to the longer range of the softer interaction potential, which decreases the cage length and obscures the intrinsic active motion. Moreover, we observe that only when the persistence length surpasses the cage length, distinct qualitative changes with respect to an equivalent passive Brownian particle system start to manifest themselves. Overall, our results further strengthen the importance of the cage length and its relation to the relevant active length scale in the context of active glassy materials.}
\end{abstract}

\maketitle %%The above information typeset through this command

\section*{Introduction}
\noindent Inspired by its ubiquitous presence in biology, active matter %systems, which typically consist of particles that can autonomously migrate through the steady consumption of energy, 
continues to be 
%have continued to be 
one of the prevalent topics in the field of biological and soft matter physics~\cite{Bechinger2016,Ramaswamy2010,Marchetti2013rev}.
Throughout the previous decade, the initial focus of active matter studies has been directed towards self-propelled particles in the dilute to moderately-dense regime~\cite{Bechinger2016}. 
%throughout the previous decade~\cite{Bechinger2016,Ramaswamy2010,Marchetti2013rev}. Within this period the initial focus has primarily been directed towards the dilute to moderately-dense regime~\cite{Bechinger2016}. 
Recently, however, active matter studies are also increasingly venturing into the high-density regime~\cite{Janssen2019active,Berthier2019review}. This has already revealed the existence of interesting and distinct non-equilibrium features such as motility induced phase separation (MIPS)~\cite{Digregorio2018,Geyer2019,Caporusso2020,Omar2021,paoluzzi2021,Cates2015}, activity-induced crystallization~\cite{Briand2016,Ni2014}, and velocity ordering~\cite{Caprini2020velocity1,Caprini2020velocity2}. By pushing the density to sufficiently large values, self-propelled systems have even been shown to reach a dynamically arrested or glassy state. Interestingly, these so-called active glassy states seem to share many characteristics with conventional passive ones~\cite{Janssen2019active,Debenedetti2001supercooled,binder2011} and have for instance been witnessed in living cells and cell layers~\cite{Zhou2009cell,Parry2014bacterial,Angelini2011cell,Nishizawa2017cell,Garcia2015cell,Grosser2021,Lama2022}, synthetic colloidal assemblies~\cite{Klongvessa2019colloid1,Klongvessa2019colloid2}, granular matter~\cite{Arora2022}, and a variety of simulations and theoretical studies~\cite{Voigtmann2017,SzamelABP2019,SzamelAOUP2015,SzamelAOUP2016,FengHou2017,BerthierABP2014,DijkstraABP2013,BerthierAOUP2017,Berthier2013activeglass,Flenner2020,FlennerAOUP2016,Henkes2011active,Reichert2020modecoupling,Reichert2020tracer,Reichert2021rev,Nandi2018,Sollich2020,janzen2021aging,Janssen2017aging,Bi2016cell,paoluzzi2021,Debets2021cage,Keta2022}.   
With the increase of density one might be tempted to automatically downgrade the role of self-propulsion on the particle dynamics in favor of the increasingly dominant particle-particle interactions. Still, even in the high-density regime active motion can affect glassy dynamics in intriguing ways~\cite{DijkstraABP2013,Berthier2013activeglass,BerthierABP2014,SzamelAOUP2015,FlennerAOUP2016,Flenner2020,Debets2021cage}. A better understanding of active glassy matter and in particular the relation to its more conventional passive counterpart has therefore opened up as a promising line of research within the broad fields of both active matter and glassy physics~\cite{Janssen2019active}. 

Importantly, it has recently been shown for quasi-hard active spheres that the cage length, i.e.\ the typical size associated with the caging of particles by their nearest neighbors, plays an important role in the context of active glassy matter~\cite{Debets2021cage,Voigtmann2017}.  Specifically, it provides a reference length to which the intrinsic short-time active length scale can be related. That is, for active length scales smaller than the cage length, dense active matter exhibits enhanced relaxation dynamics with respect to an equivalent Brownian system, while upon surpassing the cage length the relaxation dynamics starts to slowdown and eventually becomes slower than that of the passive reference system. A proposed physical mechanism underlying the observed behavior is the most efficient scanning of particle cages. This should yield the fastest relaxation dynamics and occurs when the cage length and the active length scale coincide. Consequently, the non-trivial and non-monotonic influence of activity on glassy dynamics can be understood from a conceptually relatively simple argument. How well this explanation generalizes to more complex particle-particle interactions remains, however, to be established.
%A caveat might then be that this concept proves to be too simplistic and does not generalize well to systems involving more complex particle-particle interactions. 

Here, in an effort to take a first step in the direction of more diverse interaction potentials, we demonstrate that the physical picture sketched above remains fully intact
%is in fact fully retained 
for active spheres of different softness. In short, we study the dynamics of athermal active Brownian particles (ABPs) 
%(we refrain from also considering the equally suitable active Ornstein Uhlenbeck particle (AOUP) model, since it appears that the microscopic details of these simple model systems do not significantly influence the long-time glassy behavior)
whose interactions are governed by a repulsive powerlaw potential with a variable power controlling the softness of the particles. We vary the persistence length of the constituent particles at a fixed active temperature and retrieve a qualitatively similar non-monotonic dependence of the relaxation dynamics for each considered softness. In all cases the optimum of the dynamics coincides with the point at which the persistence length is approximately equal to the cage length. We also explore the dependence of the relaxation dynamics on the active temperature upon approaching dynamical arrest, and find that the cage length marks the threshold value beyond which the active system starts behaving qualitatively distinct (manifested by changes in the fragility) from its passive Brownian counterpart. 
Note that our findings for ABPs should also apply to the equally suitable active Ornstein Uhlenbeck particle (AOUP) model, since the microscopic details of these simple model systems do not significantly influence the long-time glassy behavior~\cite{Debets2021cage}.
Overall, our work serves to further establish the importance of the cage length and in particular its relation to the short-time active length scale in the context of active glassy matter.

%the relaxation dynamics of our ABPs at persistence lengths on the order of (or below) the cage length does not exhibit any significant qualitative changes with respect to its passive counterpart.

%Such particles represent a paradigmatic model for active matter, although we mention that alternative simple active matter models could also be considered, most notably the active Ornstein Uhlenbeck particles (AOUPs). We, however, restrict ourselves to the ABP model, since it appears that the microscopic details of (at least) these simple model systems do not significantly influence the long-time glassy behavior.

\section*{Methods}
\noindent The simulation model we use is a three-dimensional (3D) Kob-Andersen binary mixture consisting of $N_{\mathrm{A}}=800$ and $N_{\mathrm{B}}=200$ athermal self-propelling soft spheres of type A and B respectively. The position $\mv{r}_{i}$ of each particle $i$ evolves in time $t$ according to \cite{Farage2015,DijkstraABP2013,FengHou2017}
\begin{equation}\label{LVeq}
    \dot{\mv{r}}_{i} = \zeta^{-1} \left( \mv{F}_{i} + \mv{f}_{i} \right) ,
\end{equation}
where $\zeta$ is the friction coefficient and $\mv{f}_{i}$ the self-propulsion force acting on particle $i$. The interaction force $\mv{F}_{i}=-\sum_{j \neq i} \nabla_{i} V_{\alpha\beta}(r_{ij})$ is derived from a repulsive powerlaw potential $V_{\alpha\beta}(r)= 4\epsilon_{\alpha\beta}\left( \frac{\sigma_{\alpha\beta}}{r}\right)^{n}$ with a variable power $n$, which controls the softness of the particles (smaller $n$ corresponds to softer particles). The interaction parameters, i.e.\ $\epsilon_{\mathrm{AA}}=1,\  \epsilon_{\mathrm{AB}}=1.5,\  \epsilon_{\mathrm{BB}}=0.5,\  \sigma_{\mathrm{AA}}=1,\ \sigma_{\mathrm{AB}}=0.8,\  \sigma_{\mathrm{BB}}=0.88$, are, in combination with setting the friction coefficient to unity $\zeta=1$, chosen to give good glass-forming mixtures~\cite{Kob1994,Michele2004}. Following the ABP model~\cite{Romanczuk2012active,Ramaswamy2017active,Lowen2020active,SzamelABP2019,Marchetti2012active,Hagen2011ABP} for our self-propulsion force, we let the absolute value of the force $f$ remain constant in time, i.e.\ $\mv{f}_{i}=f\mv{e}_{i}$, while the orientation $\mv{e}_{i}$ undergoes rotational diffusion~\cite{FengHou2017,Farage2015},
\begin{equation}
    \dot{\mv{e}}_{i} = \gv{\chi}_{i} \times \mv{e}_{i},
\end{equation}
subject to a Gaussian noise process with zero mean and variance $\avg{\gv{\chi}_{i}(t)\gv{\chi}_{j}(t^{\prime})}_{\mathrm{noise}}=2D_{\mathrm{r}}\mv{I}\delta_{ij}\delta(t-t^{\prime})$ with $D_{\mathrm{r}}$ the rotational diffusion coefficient  and $\mv{I}$ the unit matrix. In the absence of particle-particle interactions, each particle performs a persistent random walk (PRW) and its mean square displacement (MSD) is given by~\cite{FengHou2017}
\begin{equation}\label{MSDsingle}
    \avg{\delta r^{2}(t)} = 6T_{\mathrm{a}} \left(\tau_{\mathrm{p}}(e^{-t/\tau_{\mathrm{p}}} - 1) + t \right).
\end{equation}
Inspection of~\cref{MSDsingle} shows that the single-particle motion is characterized by a persistence time $\tau_{\mathrm{p}}=(2D_{\mathrm{r}})^{-1}$ and an active temperature $T_{\mathrm{a}}=f^{2}\tau_{\mathrm{p}}/3$. In particular, at short times ($t\ll \tau_{\mathrm{p}}$) the motion is ballistic $\avg{\delta r^{2}(t)}\approx 3T_{\mathrm{a}}t^{2}/\tau_{\mathrm{p}}$, and
in the long-time limit ($t\gg \tau_{\mathrm{p}}$) it becomes fully diffusive $\avg{\delta r^{2}(t)}\approx 6T_{\mathrm{a}}t$. This implies that in the limit $\tau_{\mathrm{p}}\rightarrow 0$ (with $T_{\mathrm{a}}\sim \mathrm{constant}$), our active system reduces to a Brownian one at a temperature $T$ equal to the active temperature $T_{\mathrm{a}}$. To study the effect of particle softness on the active glassy dynamics we take as our control parameters $T_{\mathrm{a}}$, the powerlaw exponent $n$, and, to quantify how far we are from the passive limit, the persistence length $l_{\mathrm{p}}=f\tau_{\mathrm{p}}$~\cite{Debets2021cage}.

Simulations are performed by solving the Langevin equation [\cref{LVeq}] via a forward Euler scheme using LAMMPS~\cite{Lammps}. We set the number density to $\rho=1.2$ via the size of the periodic simulation box, run the system sufficiently long (typically between $500$ and $20000$ time units) to prevent aging, and afterwards track the particles over time for at least twice the initialization time. Unless otherwise stated we use for the powerlaw potential a cutoff radius of $r_{\mathrm{c}}=2.5\sigma_{\alpha\beta}$. All results are presented in reduced units where $\sigma_{\mathrm{AA}}$, $\epsilon_{\mathrm{AA}}$, $\epsilon_{\mathrm{AA}}/k_{\mathrm{B}}$, and $\zeta\sigma^{2}_{\mathrm{AA}}/\epsilon_{\mathrm{AA}}$ represent the units of length, energy, temperature, and time respectively~\cite{Flenner2005}. We also mention that, to correct for diffusive center-of-mass motion, all particle positions are retrieved relative to the momentary center of mass~\cite{Flenner2005}.

\section*{Results \& Discussion}
\begin{figure*}[ht!]
    \centering
    \includegraphics [width=0.95\textwidth] {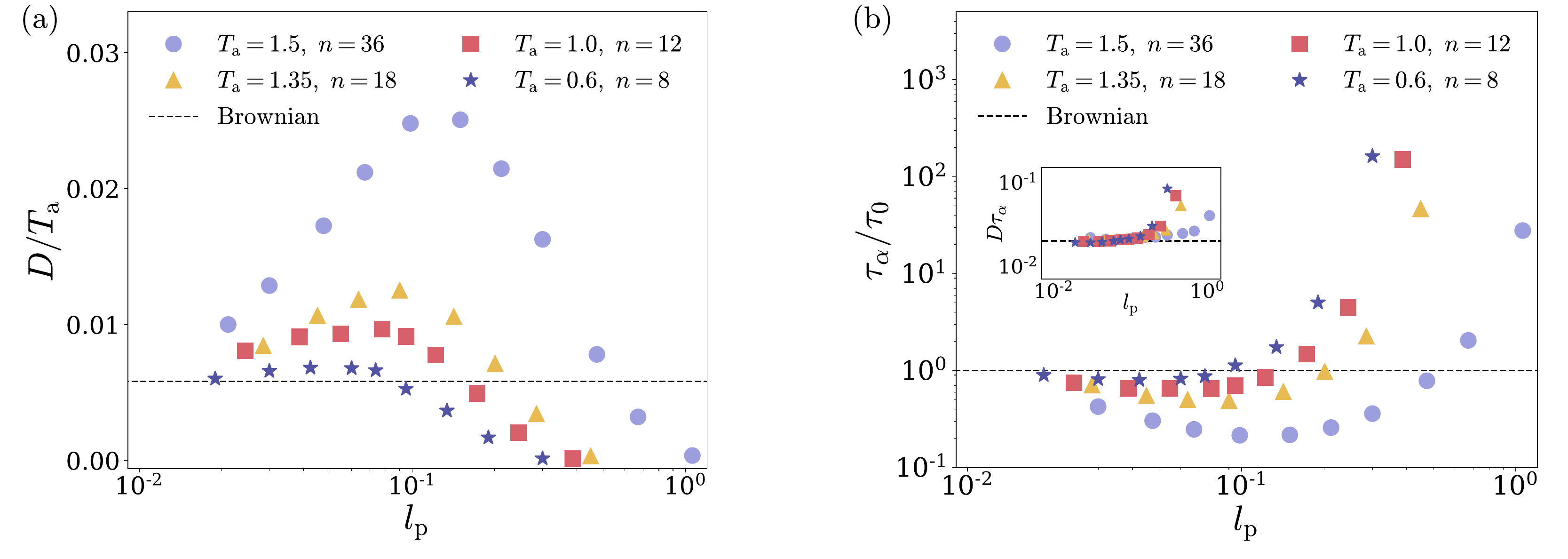} 
    \caption{The normalized (a) long-time diffusion coefficient $D/T_{\mathrm{a}}$ and (b) relaxation time $\tau_{\mathrm{\alpha}}/\tau_{0}$ as a function of the persistence length $l_{\mathrm{p}}$ for athermal self-propelling spheres with different softness (governed by the power $n$). Increasing $l_{\mathrm{p}}$ initially yields faster, but eventually slower, relaxation dynamics than Brownian particles at a temperature $T=T_{\mathrm{a}}$ (dashed line). The enhancement and optimum of the dynamics are suppressed and shifted to smaller $l_{\mathrm{p}}$ values respectively for increasing softness (smaller $n$). The inset of panel (b) denotes the Stokes-Einstein relation $D\tau_{\alpha}$, which remains constant and independent of particle softness until a threshold value of $l_{\mathrm{p}}$ on the order of the cage length is surpassed.}
    \label{Fig1}
\end{figure*}

%\noindent \textbf{Dependence on $l_{\mathrm{p}}$ qualitatively the same for different particle softness.}
\noindent In our model system the active particles become more disparate from conventional passive particles upon increasing their persistence. To understand how this relates to particle softness, we have first extracted the long-time diffusion coefficient $D=\lim_{t\to\infty}\avg{\delta r^{2}(t)}/6t$ as a function of the persistence length $l_{\mathrm{p}}$ for different powers $n$. The resulting values normalized by the active temperature $T_{\mathrm{a}}$ are plotted in \cref{Fig1}a. For each value of $n$ we have fixed $T_{\mathrm{a}}$ at a value such that the system exhibits mildly supercooled behavior and in the passive limit ($l_{\mathrm{p}}\rightarrow 0$) all different powers give the same value for the normalized diffusion coefficient. This allows for a convenient comparison. An inspection of the results shows that the qualitative shape of the curves is unaltered when increasing the particle softness. In particular, all curves demonstrate a non-monotonic dependence on $l_{\mathrm{p}}$ with initially enhanced, but eventually slower long-time diffusion than an equivalent Brownian particle at $T=T_{\mathrm{a}}$. This is consistent with previous results~\cite{Flenner2020,Debets2021cage}. Moreover, we observe that in the limit of small $l_{\mathrm{p}}$ the diffusion coefficients tend, as expected, towards Brownian dynamics result, while for large $l_{\mathrm{p}}$ they seemingly go to zero. 

Two notable quantitative differences are, however, also visible. It can be seen that the peak height strongly decreases when transitioning from quasi-hard ($n=36$) to relatively soft ($n=8$) spheres, and the location of the peak also shifts to smaller values. In previous work it has been demonstrated that the location of the peak for quasi-hard spheres corresponds to the point where the persistence length $l_{\mathrm{p}}$ is (approximately) equal to the cage length $l_{\mathrm{c}}\sim 0.1\sigma_{\mathrm{AA}}$~\cite{Debets2021cage}, i.e., the length scale of local particle caging~\cite{Hansen2013simple}. Our results seem to corroborate this claim. A physical explanation for this behavior might then be attributed to the optimal scanning of particle cages which in turn yields the fastest relaxation dynamics. Following this reasoning we believe that, at least in part, both the peak height and peak location decrease to smaller values as a result of the longer range of softer powerlaw potentials (note that we have introduced a long-range potential cutoff of $r_{\mathrm{c}}=2.5\sigma_{\alpha\beta}$). Due to the increased range, short-time particle motion becomes more perturbed by interactions so that individual soft particles cannot benefit from an efficient cage scanning as much, which explains the decreased peak height. Furthermore, the cage also becomes effectively smaller so that the optimum value coincides with a smaller persistence length. 

\begin{figure}[ht!]
    \centering
    \includegraphics [width=0.45\textwidth] {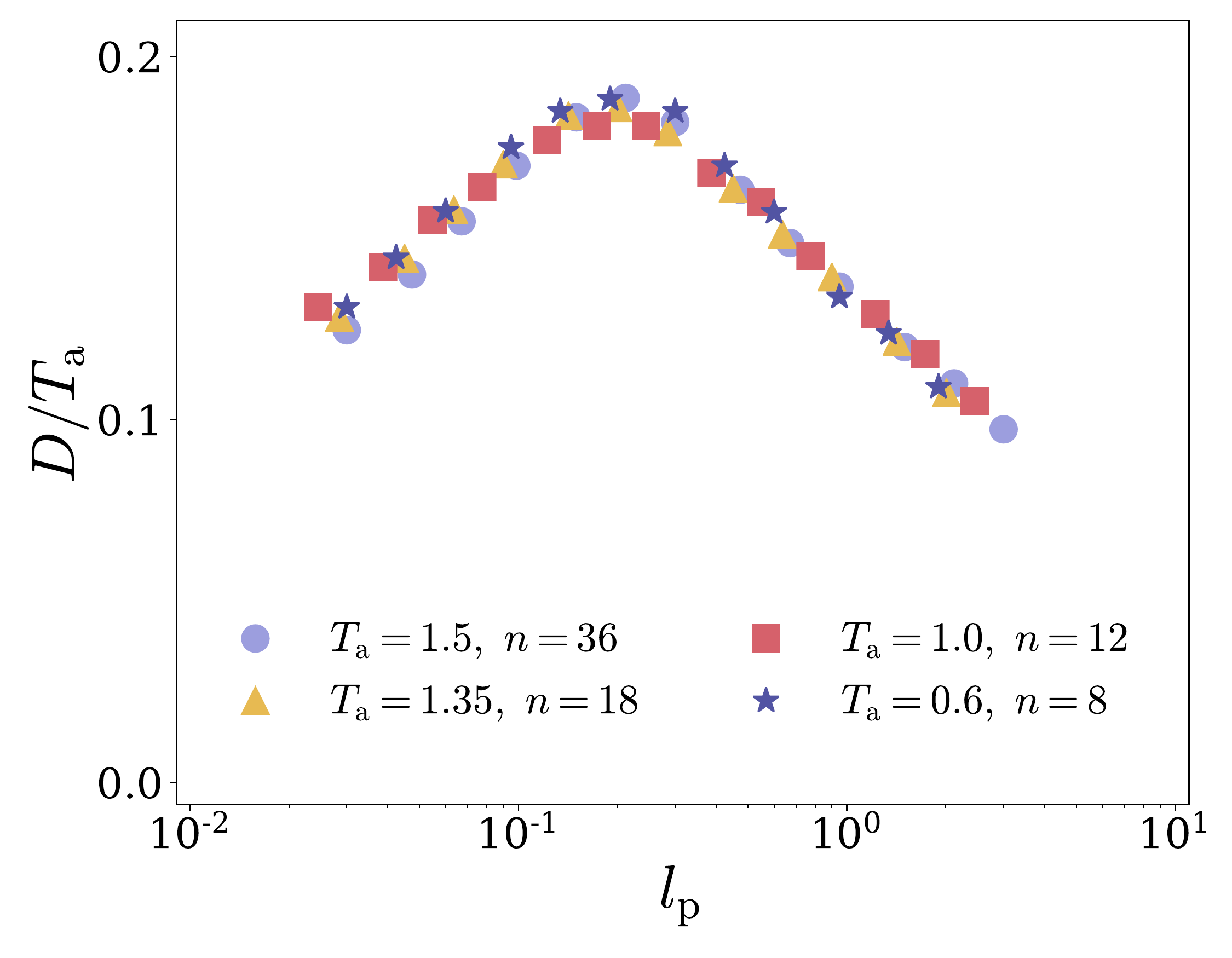} 
    \caption{Plots of the normalized long-time diffusion coefficient $D$ as a function of the persistence length $l_{\mathrm{p}}$ for different particle softness (governed by the power $n$). In comparison to the results presented in \cref{Fig1}a, the cutoff radius of the powerlaw potential is taken at a smaller value of $r_{\mathrm{c}}=1.0\sigma_{\alpha\beta}$.   }
    \label{Fig2}
\end{figure}

To test this claim, we have repeated the simulations used for the results in \cref{Fig1}a with a smaller potential cutoff radius of $r_{\mathrm{c}}=1.0\sigma_{\alpha\beta}$. In this case the range of the potential becomes shorter. It should also become less dependent on the value of $n$, since the potential immediately starts steeply increasing when the inter-particle distance becomes smaller than $r_{\mathrm{c}}$ for all considered powers, whereas this happens more gradually (especially for $n=8,12$) with a large cutoff radius. The resulting normalized long-time diffusion coefficients are shown in \cref{Fig2} where we mention that the corresponding Brownian dynamics results (at a temperature $T=T_{\mathrm{a}}$), although not exactly the same, remained of the same order ($D/T_{\mathrm{a}}\sim 0.1$) such that the results for different powers can still be accurately compared. Interestingly, the results now overlap much more and can be even seen to almost collapse. Additionally, the peak location is shifted to a larger value of $l_{\mathrm{c}}\sim 0.2\sigma_{\mathrm{AA}}$, which is still of the same order as the cage length, and seems to confirm the notion that the range of the potential determines the location of the peak. The fact that all powers now display a clear peak of approximately the same (relative) height also suggests that the steepness of the potential is an important governing factor of the peak height.

%Possibly add discussion of fitting large tau_p limit at this point
%all passive rc=1 results between 0.075 and 0.11

Next, to put our initial results (using a long-range cutoff $r_{\mathrm{c}}=2.5\sigma_{\alpha\beta}$) into a broader context we have also retrieved the self-intermediate scattering function, i.e., $F^{\mathrm{s}}(k,t)=\avg{e^{i\mv{k}\cdot \mv{r}_{j}(0)}e^{i\mv{k}\cdot \mv{r}_{j}(t)}}$, for the majority type A species. Based on these we have extracted the alpha-relaxation time $\tau_{\alpha}$, which is defined via $F^{\mathrm{s}}(k,\tau_{\alpha})=e^{-1}$ at a wavenumber $k=7.2\sigma_{\mathrm{AA}}^{-1}$ corresponding to the first peak of the static structure factor. The results for $\tau_{\alpha}$, %have in turn been 
normalized by the relaxation time $\tau_{0}$ obtained for an equivalent ($T=T_{\mathrm{a}}$) passive system,  %and 
are plotted in \cref{Fig1}b. It can be seen that the qualitative behavior of the relaxation time is fully consistent with the long-time diffusion coefficients. In particular, $\tau_{\alpha}$ initially decreases to a minimum value (indicating the fastest relaxation dynamics), which is smaller than a corresponding Brownian particle ($\tau_{\alpha}/\tau_{0}<1$), while for large $l_{\mathrm{p}}$ it increases significantly beyond this value ($\tau_{\alpha}/\tau_{0}\gg 1$). Enhanced softness again flattens the curves and shifts the optimum to a smaller value of $l_{\mathrm{p}}$. We also note that the location of the minima of $\tau_{\alpha}$ coincides with the maxima of $D$. 

To gain some insight into the influence of particle softness and persistence on our glassy model system as a whole, we have also combined the relaxation time and long-time diffusion coefficient to calculate the Stokes-Einstein relation (SER), i.e., $D\tau_{\alpha}$, which has been plotted in the inset of \cref{Fig1}b. For passive systems at large enough temperatures this relation usually remains constant, while upon vitrification significant deviations may occur~\cite{Flenner2005,voigtmann2004tagged,Tarjus2005a}. These deviations have often been attributed to the manifestation of dynamical heterogeneity, although some controversy persists~\cite{charbonneau2013dimensional}. An inspection of our results shows that the SER initially takes on approximately the same value regardless of the particle softness. The values are also similar to the ones obtained for an equivalent Brownian system ($D\tau_{\alpha}\sim 0.02\sigma_{\mathrm{AA}}^{2}$ for each considered softness), which suggests that at least for persistence lengths below the cage length the active system exhibits no distinct qualitative changes with respect to its passive counterpart. In comparison, upon further increasing $l_{\mathrm{p}}$ we observe a sudden rise of $D\tau_{\alpha}$ for all powers $n$. Interestingly, the point at which this happens seems not to concur with the optimum of the dynamics, but instead with the point at which the active dynamics becomes slower than that of the equivalent Brownian system ($\tau_{\alpha}/\tau_{0}>1$). 
Thus, the onset of slow dynamics, which is here induced by increasing the persistence length, coincides with the breakdown of the SER.
This is consistent with passive glassy phenomenology where the onset of slow dynamics is typically induced by supercooling.
%Since we are already within the mildly supercooled regime, this is consistent with the fact that for passive systems the onset of vitrification often marks the breakdown of the SER. 

\begin{figure}[ht!]
    \centering
    \includegraphics [width=0.48\textwidth] {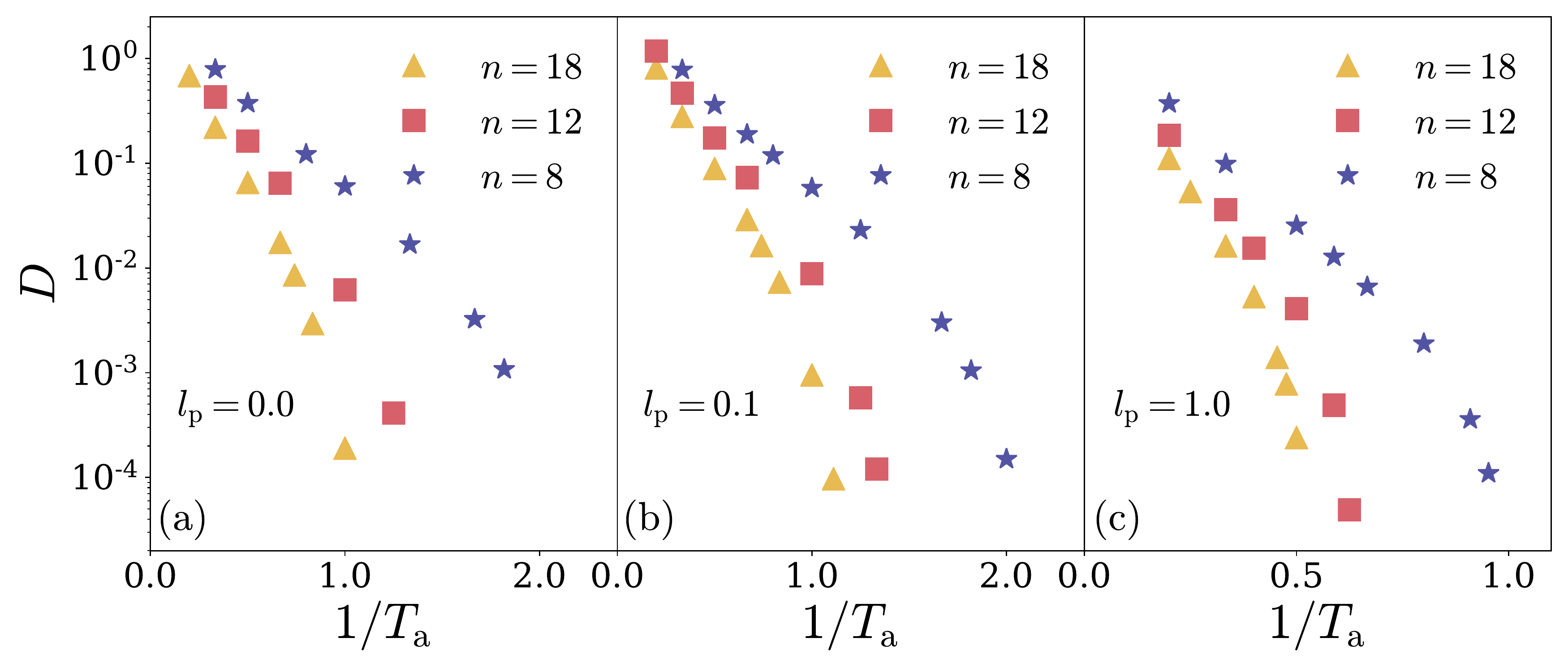} 
    \caption{Plots of the long-time diffusion coefficient $D$ as a function of the inverse active temperature $1/T_{\mathrm{a}}$ for different particle softness (governed by the power $n$) and persistence lengths $l_{\mathrm{p}}$. When the persistence length becomes larger than the cage length ($l_{\mathrm{p}}\gtrsim 0.1$), the active temperature at which $D$ tends to zero starts to significantly increase.  }
    \label{Fig3}
\end{figure}

Up until this point we have kept the active temperature fixed for each particle softness and focused primarily on the dependence on the persistence length. We now proceed by taking a more in-depth look at the qualitative and quantitative behavior of the dynamics as a function of $T_{\mathrm{a}}$. In other words, we take a closer look at how our system approaches a dynamically arrested state. Based on the non-monotonic behavior observed at a constant active temperature, we choose to concentrate on three distinct values of $l_{\mathrm{p}}=0.0,0.1,1.0$, which, in relation to the cage length $l_{\mathrm{c}}$, serve to probe the regimes $l_{\mathrm{p}}\ll l_{\mathrm{c}}$, $l_{\mathrm{p}}\sim l_{\mathrm{c}}$, and $l_{\mathrm{p}}\gg l_{\mathrm{c}}$ respectively. For these values we have calculated the long-time diffusion coefficients $D$ and plotted them as a function of $1/T_{\mathrm{a}}$ %(note that for $l_{\mathrm{p}}=0$, $T_{\mathrm{a}}$ represents the temperature $T$) 
for different particle softness $n$ in \cref{Fig3} (note that for $l_{\mathrm{p}}=0$, $T_{\mathrm{a}}$ represents the temperature $T$). We observe that in all cases the particles become slower when they are stiffer. Moreover, we see that the active temperature at which $D$ tends to zero remains approximately the same for $l_{\mathrm{p}}=0.0,0.1$, while it is significantly increased for $l_{\mathrm{p}}=1.0$. This implies that the slowdown of the dynamics when the persistence length surpasses the cage length (see for instance \cref{Fig1}) is robust for different active temperatures and is thus retained when approaching a more dynamically arrested state of our active system.  

\begin{figure*}[ht!]
    \centering
    \includegraphics [width=0.95\textwidth] {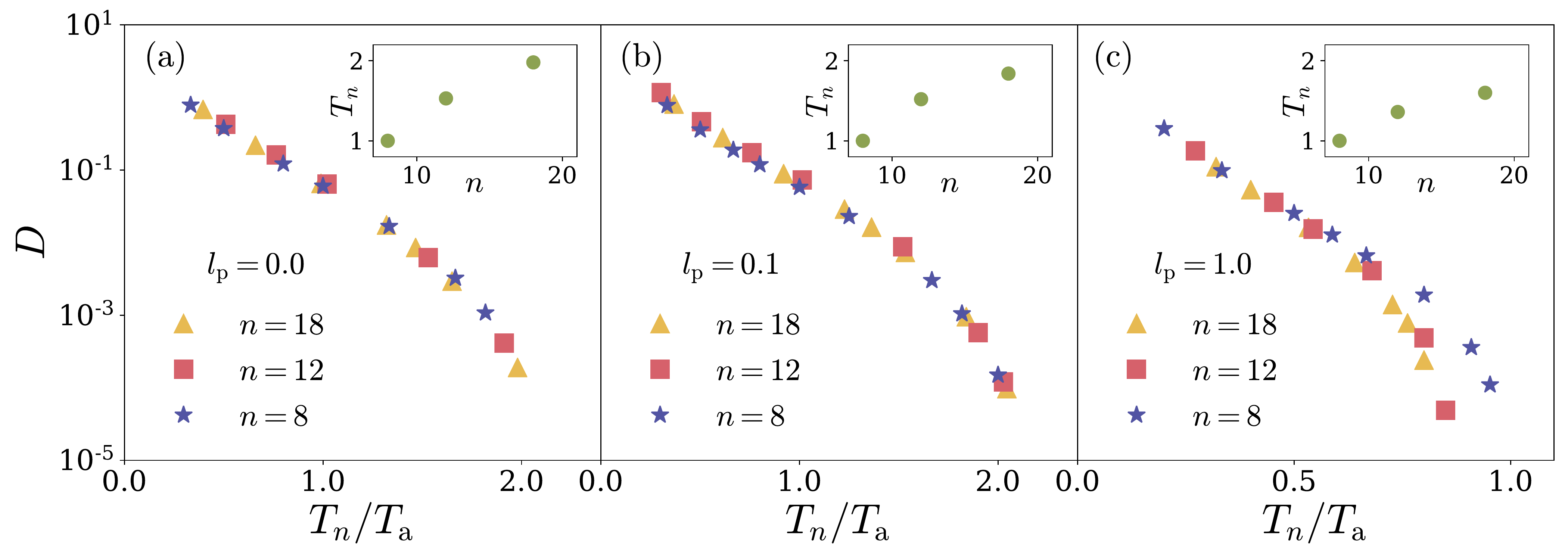} 
    \caption{Plots of the long-time diffusion coefficient $D$ as a function of the normalized inverse active temperature $T_{n}/T_{\mathrm{a}}$ for different particle softness (governed by the power $n$) and persistence lengths $l_{\mathrm{p}}$. The scaling parameter $T_{n}$ has been chosen to maximize the overlap between different curves and is plotted in the insets. For values approximately smaller than the cage length ($l_{\mathrm{p}}\lesssim 0.1$) all curves collapse onto a master curve indicating that the fragility is independent of the softness. For larger values of $l_{\mathrm{p}}$ qualitative changes occur for different particle softness.}
    \label{Fig4}
\end{figure*}

To test whether going beyond the cage length also marks the emergence of qualitative changes between different particle softness, we have sought to rescale the inverse active temperature with a scaling parameter $T_{n}$. This procedure is inspired by the fact that in previous work on a similar passive system it has been shown that (at least up to a power $n=18$) the long-time diffusion coefficients $D$ can be scaled onto a master curve (implying fragility invariance) depending solely on the scaled inverse temperature $T_{n}/T$~\cite{Michele2004}. The results of this rescaling process for our model system are demonstrated in \cref{Fig4} and indeed, in the limit of passive particles ($l_{\mathrm{p}}=0$) we also find that a scaling is possible, where we mention that the obtained values of $D$ are quantitatively consistent with the ones reported in Ref.~\cite{Michele2004}. Interestingly, such a scaling is not limited to a passive system, since our active system at $l_{\mathrm{p}}=0.1$ also exhibits a collapse of the data points. In fact, we have verified that the data for $l_{\mathrm{p}}=0.0$ and $l_{\mathrm{p}}=0.1$ can even be collapsed onto each other,  which suggests that their fragility is approximately equal and independent of $n$. On the other hand, we see that for a relatively large persistence length $l_{\mathrm{p}}=1.0$ a collapse is not possible and the fragility depends explicitly on $n$. It thus seems that only when the persistence length becomes larger than the cage length, qualitative differences with respect to a passive reference system and between different particle softness start to manifest themselves. In other words, this suggests that for all values $l_{\mathrm{p}}\lesssim l_{\mathrm{c}}$, our athermal active system can essentially be considered as a passive system with enhanced dynamics.

%We finalize our discussion by briefly studying some of these qualitative differences at $l_{\mathrm{p}}=1.0$. Specifically, we have calculated the alpha-relaxation times for the same settings as presented in \cref{Fig3}c and fitted the results using an active matter equivalent of the Vogel-Fulcher-Tamman law (VFT)~\cite{Berthier2011VFT}, i.e., $\tau_{\alpha}=C_{0}\exp[C_{1}T_{0}/(T_{\mathrm{a}}-T_{0})]$. Here, $C_{0}$, $C_{1}$, and $T_{0}$ are fit parameters. The resulting values for $\tau_{\alpha}$ and the corresponding fits are shown in fig. It can be seen that the data for each softness can be accurately fitted with the VFT-law. Moreover, in concurrence with the diffusion coefficient $D$, stiffer particles (larger $n$) take longer to relax. From the fit parameter $C_{1}$ we may then obtain an estimate of the fragility. In particular, its values are $C_{1}=$ for $n=8,12,18$ respectively. Since in the context of passive particles a smaller value for $C_{1}$ is typically associated with a more fragile glass~\cite{Berthier2011VFT}, we conclude that upon moving past the cage length softer particles seem to become more fragile. 

\section*{Conclusion}
\noindent In this work we have, by means of extensive computer simulations of athermal active Brownian particles (ABPs), explored the subtle relationship between active motion and particle softness in the glassy regime. Our results demonstrate that the qualitative behavior of the relaxation dynamics at a fixed active temperature is robust to changes in softness. In particular, the relaxation dynamics exhibits a non-monotonic dependence on the persistence length (the intrinsic active length scale) with an optimum (largest speedup) corresponding to the point where the persistence length coincides approximately with the cage length. Small quantitative differences for varying softness have in turn been rationalized by considering the longer range of the softer interaction potential, which decreases the cage length and obscures the intrinsic active motion. As a result, the optimum of the dynamics shifts to smaller persistence lengths and becomes flattened for increasing softness. 

%This optimum is shown to decrease in size, while its location shifts to smaller values of the persistence length for softer particles. We have explained these quantitative differences by relating them to the longer range of the softer interaction potential, which decreases the cage length and obscures the intrinsic active motion. 

When the persistence length is instead kept fixed at a value approximately equal to the cage length we witness the appearance of a universal curve (see \cref{Fig4}b) onto which the long-time diffusion coefficients for different softness (up to a power $n=18$) fully collapse as a function of the scaled active temperature. In fact, even the results of an equivalent Brownian system can be added to this universal curve (see \cref{Fig4}a). Consequently, the relaxation dynamics of the active system at relatively small persistence lengths does not exhibit any significant qualitative changes with respect to its passive counterpart and its fragility is independent of the softness.
In contrast, when the persistence length is set at a significantly larger value than the cage length we find that both the active temperature at which the long-time diffusion coefficient tends to zero starts to significantly increase and a collapse of the long-time diffusion coefficients is no longer possible. The latter indicates that the qualitative features of vitrification, e.g., the fragility, explicitly depend on particle softness.
%and distinct behavior for different particle softness and with respect to passive particles emerges. This manifests itself via a larger critical active temperature and changes in fragility. 
%and a seemingly smaller fragility for softer particles.
%which implies that the qualitative features of vitrification explicitly depend on particle softness. In particular, we find that the softer particles becomes less fragile 
%As a result, the active temperature at which $D$ tends to zero starts to significantly increase with respect to a passive Brownian system 

Overall, our work shows that the cage length marks the offset beyond which active glassy matter becomes qualitatively different from conventional passive glassy materials. 
It therefore further strengthens the importance of the cage length and its relation to the relevant active length scale in the context of active glassy dynamics. As a followup it might be worthwhile to check the role of the cage length for model biological glass-formers such as confluent cell layers~\cite{Janssen2019active,Bi2016cell,Angelini2011cell,Ruscher2020} or for (colloidal) systems involving more complex and possibly attractive interaction potentials. Alternatively, a more detailed study on the qualitative changes for different softness in the limit of large persistence~\cite{Keta2022,Mandal_2021} or an analysis of our results in relation to a recently introduced mean-field softness~\cite{Nandi2021softness} are equally interesting, but for now left for future work.

Finally, we want to mention that in recent years much work has also been devoted to the escape properties of a single active particle from a potential trap or within a porous environment~\cite{Kurzthaler2021,Wexler2020,Militaru2021,Woillez2019,Caprini2021,Chaudhuri_2021,Malakar2020}. It would be interesting to check whether the analogy between a potential trap or porous environment and a dynamic cage of surrounding particles can be exploited to better understand the qualitative features, particularly the non-monotonic behavior, of dense active matter.   

\section*{Acknowledgments}
\noindent We acknowledge the Dutch Research Council (NWO) for financial support through a START-UP grant (V.E.D. and L.M.C.J.).

%\section*{SUPPLEMENTARY MATERIAL}

\bibliographystyle{apsrev4-1}
\bibliography{all}

% Figure legends
%%Automatically it will add the figure legends  and table legends as a list by below command

%\newpage

%\listoffigures

%\newpage

%\listoftables

% Figures and Tables coding should be placed where the
% first reference in the text.
% All the Figure files should be placed same working directory,
% for example (fig_1.eps and fig_1.pdf files must be present
% in the document directory)

% closing statement, nothing below matters

\end{document}